\documentstyle[psfig,referee]{mn}

\newcommand{\etal}{{\it et al.\,}}

\newcommand{\lapprox}{\lower0.8ex\hbox{$\buildrel <\over\sim$}}
\newcommand{\gapprox}{\lower0.8ex\hbox{$\buildrel >\over\sim$}}
\newcommand{\ergs}{\mbox{${\,\rm erg~s}^{-1}$\,}}

\def\etal{{\it et al. }}

\newcommand{\kt}{\mbox{$kT$}}

\newcommand{\rosat}{\mbox{\sl ROSAT}}

\newcommand{\einstein}{\mbox{\sl EINSTEIN}}
\newcommand{\axaf}{\mbox{\sl Chandra}}
\newcommand{\ginga}{\mbox{\sl Ginga}}

\title[A Multiwavelength View of NGC\,253]{A Multiwavelength View at the
Heart of the Superwind in NGC\,253}

\author[D. A. Forbes et al.]
{Duncan A. Forbes$^1$, Edward Polehampton$^1$, 
Ian R. Stevens$^1$, Jean P. Brodie$^2$ \cr
Martin J. Ward$^3$\\ 
$^1$ School of Physics and Astronomy, University of Birmingham, 
Edgbaston, Birmingham, B15 2TT, UK\\
$^2$ Lick Observatory, University of California, Santa Cruz, CA 95064,
USA\\
$^3$ X--ray Astronomy Group, University of Leicester,
Leicester LE1 7RH, UK}

\date{Accepted .....................; Received .....................; 
in original form .......................}

\begin{document}

\maketitle

\begin{abstract}

Although NGC 253 is one of the nearest starburst galaxies, the activity in
its central regions is yet to be fully understood.
Here we present new optical data from the {\it Hubble Space Telescope} of
its central region, which reveal numerous discrete sources in a
ring--like structure. This is combined with data at infrared,
millimeter, radio and X--ray wavelengths to examine the nature of these
discrete sources and the nucleus itself. 

We find that the majority of optical/IR/mm sources are young
star clusters which trace out a $\sim$ 50 pc ring, 
that defines the inner edge of a cold gas
torus. This reservoir of cold gas 
has probably been created by gas inflow from a larger scale bar and
deposited at the inner Lindblad resonance. The family of compact radio
sources lie interior to the starburst ring, and in general
do not have optical or IR counterparts. They are mostly SNRs. 
The radio nucleus, which is probably an AGN, 
lies near the centre of the ring. The X--ray
emission from the nuclear source is extended in the \rosat ~HRI detector
indicating that not all of the X--ray emission can be associated with the AGN. 
The lack of X--ray variability and the flat radio spectrum of the nucleus,
argues against an ultraluminous SN as the dominant energetic source at the
galaxy core. 
The diffuse emission associated with the outflowing superwind is present in
the central region on a size scale consistent with the idea of collimation
by the gas torus.

\end{abstract}


\section{Introduction}

The central regions of galaxies are often the focus of extragalactic
studies. It is here that stellar and gas
densities are the highest, which in turn may lead to an intense
starburst or AGN activity associated
with a black hole. Although intensively studied, our understanding of
the central activity in galaxies is still
somewhat limited. For example, what is the relationship between
an AGN, a circumnuclear starburst and an outflowing superwind ?

Within a few Mpc of the Milky Way there are three starburst galaxies,
with superwind emission.
They are NGC\,253 in the Sculptor group of galaxies,
M82 in the M81 group and NGC\,1569. Both 
NGC\,253 and M82 galaxies are highly inclined spirals,
revealing a strong nuclear starburst and
an outflowing superwind perpendicular to the disk, while NGC\,1569 is a
dwarf starburst (Heckman \etal 1995). 
Both M82 and NGC\,253  may
harbor massive black holes (see Wills
\etal 1999 and Ptak \& Griffiths 1999 for evidence at radio and X--ray
wavelengths for evidence for an AGN in M82), while for
NGC\,1569 there is no real evidence for a massive black hole. 

As part of this study, we will discuss the evidence for
a massive black hole in NGC\,253. Here we 
present new narrow band images
of the central regions of NGC 253 
using the {\it Hubble Space Telescope} (HST). At a
distance of 2.5 Mpc, the $\sim$ 0.1$^{''}$ resolution of HST
corresponds to 1.21 pc, thus allowing us to probe central star formation
on size scales relevant to individual star clusters. 
We also analyse X--ray images of
NGC\,253 
from the \rosat\ database and include data from the
literature at other wavelengths. 
Although NGC\,253 is a remarkably complex system, with a very extensive
and interesting superwind, in this paper we focus on the central
regions, using optical, infrared, mm, radio and X--ray 
observations to study the
processes right at the heart of the superwind.

\section{Observations and Data Analysis}

\subsection{HST [SIII]9532 images}

We obtained new observations of the central region of the starburst
galaxy NGC\,253 on 1997 July 9 using the Wide Field and Planetary Camera 2 
(WFPC2) of the {\it Hubble Space Telescope}.  The pointing was such that
the galaxy nucleus was centred in PC chip and the WFC chips imaged the
region SE of the nucleus. Two exposures were taken through each of the F953N 
($\lambda$$_{c}$$\backslash$$\Delta$$\lambda$ = 9545$\backslash$53 \AA) 
and F850LP ($\lambda$$_{c}$$\backslash$$\Delta$$\lambda$ =
9101$\backslash$986 \AA) filters of 1300s and 300s respectively. These
correspond to the line emission of [SIII]9532 and nearby continuum. 

For each of the two filters we found no significant shift 
between the two exposures and so combined them using the STSDAS task
{\it crrej}. This effectively removed cosmic rays. 
Each image was then scaled to an 
exposure time of 1 second. 
The continuum image was scaled for filter throughput, based on the WFPC2
handbook, and then subtracted from the narrow band image to produce a
line--only [SIII]9532 image. The final image is shown in Fig. 1.

We also obtained WFPC2 archive data on the nuclear region of NGC\,253.
Images through filters F656N 
($\lambda$$_{c}$$\backslash$$\Delta$$\lambda$ = 6564$\backslash$21 \AA) 
and F675W ($\lambda$$_{c}$$\backslash$$\Delta$$\lambda$ = 6726$\backslash$866
\AA) covering a similar spatial region as our data were available. 
The exposure times were 2 $\times$ 1200s and 2 $\times$ 200s respectively. 
In this case the galaxy nucleus was centred on WF3. 
These images were then reduced in a similar way to the F953N and F850LP
data.  
The final line--only H$\alpha$ image contains one foreground star, 
which allowed
us to check our continuum subtraction in this case.\\


\subsection{\rosat\ HRI and PSPC images}

Observations of NGC\,253 have been carried out with both the HRI and PSPC 
instruments onboard \rosat. We obtained three HRI and
two PSPC  observations from the Leicester Data Archive. Details of the
observations are given in Table 1. 

The data were reduced and analysed using the Starlink 
{\it ASTERIX} software. 
The PSPC data were cleaned of periods of high background and poor pointing
stability, leaving a total of 22,015s of good data between 
the two images. The data were sorted to form a spectral image over a 
0.85$^{\circ}$ $\times$ 0.85$^{\circ}$ region centred on the nucleus 
of NGC\,253. A model of the background was constructed using data from an 
annulus {\it r}=0.2$^{\circ}$ to {\it r}=0.3$^{\circ}$. The task 
{\em pss} was used in 
several iterations to 
remove sources from the background annulus which was then scaled up to the
whole image size allowing for energy dependant vignetting. Source positions 
in each image were then measured and the two images were found to have no
significant shift. They were combined using the task {\em add}. 
The background files for the two 
images were also combined. Finally the {\em pss} task was used to locate and
measure point sources. The PSPC image is shown overlaid on an optical image
of NGC\,253 in Fig. 2. The main features are the extended
superwind emission extending about 12$^{'}$ (9 kpc) above the plane of the
galaxy with emission also extending
along the disk of the galaxy. There are a large number of X-ray
point sources, both in the plane of the galaxy and in the near
vicinity. The superwind emission has been studied before by 
Read \etal (1997), Fabbiano (1988) and Dahlem, Weaver \& Heckman (1998). 
The superwind X-ray 
emission is generally similar in structure to that seen
in M82 (Strickland, Ponman \& Stevens 1997). In this paper, we 
focus on the inner regions of the superwind, and will not discuss
the extended emission further. The X--ray data on the inner regions and
the X--ray point sources there will be discussed in more
detail in Section~3.3.

For the HRI data, the three data sets were sorted into images of size 
0.3$^{\circ}$ $\times$ 0.3$^{\circ}$ centred on the nucleus 
and then the positions of bright sources determined. 
The background was calculated by taking the mean value of the image after
the bright sources had been removed. No vignetting correction was applied. 
The images were then aligned and combined, giving a 
total effective exposure time of 56,494s. The HRI image of the
central regions is shown in Fig. 3 (left panel), and 
the HRI image overlaid on the 
HST H$\alpha$ image is shown in Fig. 3 (right panel).

Both the HRI and PSPC images clearly reveal a number of point
sources in the disk/bulge but also diffuse disk emission and a spectacular
bi--conical emission perpendicular to the disk in the case of the PSPC (see
Fig. 2.



\section{Discrete Sources}

\subsection{Optical, Infrared and Millimeter Sources}

Using a then new 256 $\times$ 256 IR array  and high spatial sampling
(0.05$^{''}$/pixel in 0.7$^{''}$ seeing conditions) Forbes \etal (1991)
detected several `hotspots' or discrete sources in the central region of
NGC\,253 at 1.65$\mu$m (H band). Since then several other imaging studies
of NGC\,253 have been carried out, e.g. Br$\gamma$, [FeII], H$_2$, 
K band (Forbes \etal
1993); 8.5, 10, 12.5$\mu$m (Keto \etal 1993); 10$\mu$m (Pina \etal 1993);
J, H, K bands (Sams \etal 1994); H, K, L, M bands and PAH feature 
(Kalas \& Wynn--Williams 1994); [NeII] and 12.8$\mu$m (Boker \etal 1998).  
These studies have also revealed multiple discrete sources. 

Measuring absolute positions for these sources has been problematic given
the small field-of-view of the IR arrays.
Forbes \etal (1991) assumed that the brightest 2.2$\mu$m (K band) source
was at the dynamical centre 
and therefore spatially coincident with the radio nucleus
(source 2 in Turner \& Ho 1985). More recent astrometry has shown that this is
{\it not} the case, and that the IR maximum is actually $\sim2^{''}$W and
$\sim1^{''}$S of the radio nucleus (Kalas \& Wynn--Williams 1994). The position
of this IR maximum is constant from 1.6 to
4.8$\mu$m (Kalas \& Wynn--Williams 1994) and even to 12.5$\mu$m (Keto \etal
1993). Taking a simple mean of the position from Kalas \& 
Wynn--Williams (1994),
Keto \etal (1993) and Pina \etal (1992) we calculate its location 
to be $\alpha$ = 00$^{h}$47$^{m}$33.006$^{s}$ $\pm 0.01^s$, $\delta$ 
= --25$^{\circ}$17$^{'}$18.23$^{''}$ $\pm 0.2^{''}$
(J2000). Thus the brightest IR source 
is assumed to lie at this position and all 
other IR sources are shifted relative to it. 

The brightest source in the [SIII]9532 image is located at 
$\alpha$ = 00$^{h}$47$^{m}$32.968$^{s}$, 
$\delta$ = --25$^{\circ}$17$^{'}$18.47$^{''}$ (J2000) according 
to the STSDAS astrometry task
called {\it metric}. This position is within 0.51$^{''}$ and 0.24$^{''}$ in
R.A. and Dec. of the IR maximum. 
According to the HST handbook, the {\it metric} derived positions
could have an absolute positional error of $\pm 0.5^{''}$. 
We will assume that the brightest [SIII] source is spatially coincident with
the IR maximum and shift it by $\Delta$R.A. = 0.51$^{''}$ and
$\Delta$Dec. = 0.24$^{''}$. 
Given the extinction in the central region, 
this assumption may be questionable. However, as we show later, 
some of the other [SIII] sources reveal a good positional agreement with
other IR sources after the offset. This gives us confidence that the
brightest [SIII] source is indeed associated with the brightest IR source. 

The H$\alpha$ image 
has a very similar morphology to the [SIII] image, so there is
little chance of confusion in comparing the images. We find that the
brightest [SIII] source corresponds to the second brightest H$\alpha$
source.  
We have shifted this source to match the position of the IR maximum 
as well. 

The brightest 11 discrete sources in the [SIII]9532 and H$\alpha$ images
form an elliptical ring (see Fig. 1). 
The ring has rough dimensions of $\sim 4^{''}$ (48
pc) in the SW--NE direction and $\sim 1.5^{''}$ (18 pc) in the SE--NW
direction.  The position angle is $\sim$ 60$^o$. The projected ellipticity
of the ring is about 0.6, however if it lies at an inclination of 
78$^o$ (i.e same as the overall inclination of
the galaxy), then it is intrinsically almost circular.  
Beyond the ring there are 4 clearly identifiable
discrete sources. We have measured the counts in a 0.4$^{''}$ diameter
aperture centred on each source maximum. These have been converted into
flux using equation 11 from Holtzman \etal (1995). In Table 2 we list the
corrected positions,
offset from the radio nucleus 
(i.e. $\alpha$ = 00$^{h}$47$^{m}$33.169$^{s}$, 
$\delta$ = --25$^{\circ}$17$^{'}$17.06$^{''}$ J2000), the [SIII]9532 flux,
the H$\alpha$ flux and the dereddened ratio of [SIII]9532+9069 to
H$\alpha$.  

The WFPC2 H$\alpha$ image has been published previously by Watson \etal
(1996). They also described the H$\alpha$ sources as forming a ring--like
structure but only measured the flux from 4 of them. Confusingly the
`bright blob' they refer to is the brightest continuum source, but only
ranks as the second brightest H$\alpha$ line emission 
source. Like us, they used 
0.4$^{''}$ diameter apertures.  
Rather than use the Holtzman \etal (1995) formula
they compared the count rates from each source with an  
assumed power--law model for the continuum to get the required flux
levels. In units of $10^{-15}$ erg
cm$^{-2}$ s$^{-1}$, their measurements followed by ours in brackets are 6.8
(1.5), 3.5 (0.8), 5.3 (1.0), 1.0 (0.3). Thus the Watson \etal 
values are typically a factor of 3 greater. This discrepancy could be due
to different centerings and/or a different counts-to-flux
conversion, but without further details of the Watson \etal method we can
only speculate. 

The central regions of NGC\,253 have also been observed in the HCN $J$ = 1
$\rightarrow$ 0 transition (Paglione, Tosaki \& Jackson 1995). 
This molecular line traces dense ($n > 10^4$ cm$^{-3}$) gas. They found
that the HCN--emitting clouds lie along a line in the SW to NE direction. 
The brightest, and highest density, HCN 
source lies 0.5$^{''}$ in R.A. and 2$^{''}$ in Dec. from the IR maximum. 
The beam
size used was 4.2 $\times$ 2.2$^{''}$, with a claimed positional accuracy
of $\pm$0.4$^{''}$. It is debatable whether the brightest HCN source is
spatially coincident with the IR maximum. 
We have decided that is probably is, and
have shifted each of the other 7 HCN sources appropriately. In the
discussion below, the reader should bear in mind the
possibility that this HCN source is not actually associated with the IR 
maximum. The other 7
HCN sources lie beyond the ring region. 
The corrected positions of the infrared and mm sources are summarised in
Table 3.

\subsection{Radio Sources}

High spatial resolution radio maps of NGC\,253 have been obtained at 
1.3, 2, 3.6, 6
and 20cm by Turner \& Ho (1985), Antonucci \& Ulvestad (1988) and  
Ulvestad \& Antonucci (1991, 1994, 1997). 
These maps reveal a family of discrete
sources. The position of these sources within different maps agree to about
$\pm 0.1^{''}$. We have decided to adopt the 2cm mapping by Ulvestad \&
Antonucci (1997) because of its high spatial resolution and good
sensitivity. The brightest radio source (about three times brighter than 
the next
strongest) which has a flat spectrum ($\alpha^6_2 = +0.04 \pm 0.06$)
and high brightness temperature (T$_B \sim 10^5$K), 
appears to be a compact
synchrotron source. This source (also
known as TH2 from Turner \& Ho 1985) is  probably an AGN and the true
nucleus of NGC\,253. We will refer to this source as the `radio nucleus'. 
It is located at 
$\alpha$ = 00$^{h}$47$^{m}$33.169$^{s}$, 
$\delta$ = --25$^{\circ}$17$^{'}$17.06$^{''}$ (J2000). 
The positions, 2cm fluxes, 2cm luminosities and 6 to 2cm spectral
indices for the 16 strongest 2cm radio sources are listed in Table 4. 

In Fig. 4 we show the location of the H$\alpha$, 
[SIII]9532, IR, mm and radio discrete sources seen in the central $\sim$
50pc after aligning the different sources as described above. There is a
good spatial correspondence between the H$\alpha$ and [SIII] sources, and
to some extent with the IR sources. The elliptical 
ring--like structure can be clearly seen, with the nucleus close to the
centre of the ring. On the other hand the radio sources do {\it not} 
show a good
spatial correspondence to the ring sources and generally trace out a line
in the SW--NE direction interior to the ring. In particular, 
the radio nucleus does not have a strong IR counterpart (e.g. Sams \etal
1994) and lies in an empty region of the [SIII] emission line ring.


\subsection{X--ray Sources} 

The X-ray emission from NGC\,253 has been studied previously. Fabbiano
(1988) reported on \einstein\ IPC observations, which revealed a number
of point sources embedded in the disk of the galaxy, as well as extended
superwind emission. 
\ginga\ observations, discussed by Ohashi \etal (1990) suggested 
that the bulk of the X--ray emission from NGC\,253 is thermal (i.e. from
a superwind or individual SNRs) and that other sources make only a small
contribution. 

Ptak \etal (1997) reported on {\sl ASCA} observations of both M82 and
NGC\,253. The {\sl ASCA} satellite has much poorer spatial resolution
compared to the \rosat\ observations presented here, but better
spectral capabilities. Data for NGC\,253 were extracted from a region
of 6$^{'}$ radius. The spectra from this region was fitted with a two
component model, with the soft component being thermal with $\kt\sim
0.8$~keV and the harder component being either thermal with $\kt\sim 7$~keV
or a powerlaw with $\Gamma=2.0$. 
Dahlem \etal (1998) studied NGC\,253 along with six other 
edge--on starburst galaxies. Combining spectroscopy from 
\rosat\  and {\sl ASCA} they estimated the temperature for a soft and
medium component to be  $\kt\sim 0.27$ and 0.72~keV respectively. The hard
component had a similar powerlaw slope as found by Ptak \etal .
{\sl BeppoSax} observations of NGC\,253, presented by Mariani \etal (1999),
found that the harder spectral component was likely thermal and found
evidence for line emission at 6.7~keV. 
Data from \rosat\ observations of NGC\,253 have been presented by
Read \etal (1997) in their survey of X-ray emission from
nearby spiral galaxies.

Read \etal (1997) found a total of 15 point
sources within a radius of 15${'}$ of the galaxy center, and that the
diffuse emission comprises 74\% of the total emission.
The X--ray point sources seen in the \rosat\ data have recently been
analysed in detail by Vogler \& Pietsch (1999). They catalogued 
a total of 73 X-ray point sources, 32 of
which are associated with the galaxy disk (see also Pietsch 1992 for a
preliminary analysis  of the \rosat\ data). Most, perhaps all, of
these point sources are X--ray binaries. A similar conclusion was
reached in the earlier study of Fabbiano \& Trinchieri (1984) using the
\einstein\ satellite. 
For simplicity, we adopt the naming convention of Vogler \& Pietsch
(1999) for the point sources.

In this paper we are concerned with the 
inner $\sim$1 kpc region, contains a central source (labelled X34 in
Vogler \& Pietsch), a point--like source (X33), possibly a third
source ($\alpha$ = 00$^h$47$^m$34.3$^s$, $\delta$ = 
--25$^{\circ}$17$^{'}$50$^{''}$) and surrounding diffuse emission (see 
Fig. 3). 

The two sources (X33 and X34) are separated by 27$^{''}$, which is
comparable to the resolution of the PSPC but reasonably well separated by the HRI. Thus we are confident about 
obtaining accurate HRI--derived luminosities (given an assumed spectral
shape) but attempts to obtain spectral fits of the two sources with the
PSPC may be contaminated by emission from the other source, as well as
diffuse emission. 

Starting with the point source X33 in the HRI data, we assume the same 
spectral parameters as Fabbiano \& Trinchieri (1984) and Vogler \&
Pietsch (1999), i.e. 
Galactic hydrogen column density of $N_H = 1.7 \times 10^{20}$ cm$^{-2}$
and thermal bremsstrahlung spectrum of $\kt = 5$keV. We
derive an emitted 0.2--4.0keV flux of $3.44 \times 10^{-13}$ erg cm$^{-2}$
s$^{-1}$ and corresponding luminosity of $L_X=2.57\times 10^{38}\ergs$. 
This is in reasonable agreement with Vogler \& Pietsch value for
X33 of $2.76\times 10^{38}\ergs$ (when adjusted to our assumed
distance).  Based on its high luminosity, evidence for variability and
possible hard spectrum (from the PSPC data), Vogler \& Pietsch conclude
that it is a black hole X--ray binary. A similar conclusion was reached by 
Fabbiano \& Trinchieri (1984). We will not discuss this source further, and
concentrate on the nuclear source and surrounding diffuse emission. 

The nuclear X--ray source in the HRI image occurs at 
$\alpha$ = 00$^{h}$47$^{m}$33.32$^{s}$, 
$\delta$ = --25$^{\circ}$ 17$^{'}$ 23.0$^{''}$ (J2000) with a pointing error of
about $\pm 8.5^{''}$. For the PSPC, it is at 
$\alpha$ = 00$^{h}$47$^{m}$33.33$^{s}$, 
$\delta$ = --25$^{\circ}$ 17$^{'}$ 27.4$^{''}$ $\pm 20^{''}$.
The equivalent \einstein\ source is located at 
$\alpha$ = 00$^{h}$47$^{m}$ 33.4$^{s}$, $\delta$ = 
--25$^{\circ}$17$^{'}$23.0$^{''}$ $\pm 5^{''}$. All three sources 
lie at the centre of more extended X--ray emission, and their positions 
agree within the
uncertainties. They lie close to the position of the radio nucleus. 
We will therefore assume that they are the same source 
and that the peak of the X--ray emission is associated with the radio 
nucleus, i.e.  
$\alpha$ = 00$^{h}$47$^{m}$ 33.169$^{s}$, $\delta$ = 
--25$^{\circ}$17$^{'}$17.06$^{''}$ (J2000). 

The nuclear source has a FWHM size of about 10$^{''}$ and is clearly 
extended in the HRI image. 
This indicates that not all of the emission can
be associated with an AGN. 
We estimate an emitted 0.2--4.0 keV 
flux from the nuclear
source to be  $1.84\times 10^{-11}$ erg cm$^{-2}$ s$^{-1}$ for an internal
extinction towards the nucleus of A$_V$ = 10$^m$ (Rieke \etal
1980). This gives a 
luminosity of $L_X=1.37\times 10^{40}\ergs$. The ASCA observations
described by Ptak \etal (1997, 1999) suggest a hard component with a
luminosity of $L_X \sim 4 \times 10^{39}\ergs$. 

We note for comparison that the analysis of \rosat\ PSPC data by Read
\etal (1997) determined the following spectral properties for the
central source, \kt=3.0keV  and $N_H=6.05\times 10^{20}$ cm$^{-2}$ for a
single temperature bremsstrahlung model. Note, that the PSPC data includes
contributions from source X33 as well as the nuclear source X34.

\section{Nature of the Discrete Sources}

\subsection{The Nucleus}

As mentioned above, the nuclear radio source (known as TH2) has 
a flat radio spectrum and high brightness temperature. It has 
a compact ($<$ 2pc) core
(Sadler \etal 1995) and associated
H$_2$O maser emission (Nakai \etal 1995). These facts argue strongly for an
AGN at the centre of the galaxy.

We estimate the 
intrinsic soft X--ray emission from the nuclear source to be $\sim$ 
10$^{40}$ erg s$^{-1}$ assuming an extinction of A$_V$ = 10$^{m}$ (Rieke \etal
1980). Vogler \& Pietsch (1999) estimate 10$^{39}$ erg s$^{-1}$ for an
internal column density of $N_H= 2-3\times 10^{21}$ cm$^{-2}$, based
on a spectral fit of the PSPC data. As mentioned earlier, such fits should
be taken with caution as the bright X--ray binary to the South of the
nuclear source contaminates the PSPC flux. Another reason to be skeptical of
this result is that the column density implies an extinction of A$_V$ =
1.3$^m$. If an AGN were responsible for the X--ray emission and hidden by only
1.3 mags of extinction, then we would expect to see evidence for a Seyfert
nucleus in near--IR spectra (A$_K$ $\sim$ 0.13) which has not been
observed. Indeed, 
from high resolution near--IR imaging, Sams \etal (1994) 
have suggested that the extinction towards the nucleus could be in excess
of A$_V$ = 24$^m$. In which case, even our estimate of  
the X--ray luminosity is severely underestimated. 

The high X--ray luminosity, and spatial extent, effectively rules out 
an X--ray binary as the source of the X--ray emission. A single
Galactic--like SNR is also ruled out. There is good evidence from the radio
data for an AGN, however the extended X--ray emission implies that some of
the X--rays are {\it not} from an AGN but perhaps from a 
small number of compact, extra--luminous SNRs or hot gas associated with
starburst activity. The satellite \axaf, with its 0.5$^{''}$ spatial
resolution, should settle this issue. 

For comparison, we note that the X--ray luminous radio supernova
SN\,1988Z, has radio and X--ray properties that are broadly
comparable to the nucleus of NGC\,253. SN\,1988Z was detected as an
X--ray source with $L_X\sim 10^{41}\ergs$, at an epoch $\sim 8$ years
after the SN event (Fabian \& Terlevich 1996). It has also been
studied at radio wavelengths, where it is a luminous (but fading)
object (van Dyk \etal 1993). The overall spectral shape consists of a
power--law with slope $\alpha=-0.74$ and varying free--free absorption.
Another similar object is SN\,1986J, which is again a radio and X--ray
luminous object (Weiler \etal 1990, Houck \etal 1998).

The peculiar X--ray and radio characteristics of these objects have been
interpreted as being due to the SN explosion occurring in a dense
circumstellar environment, caused by an earlier phase of stellar
evolution (e.g. dense winds) or location in a
dense environment (e.g. molecular cloud or starburst region). The implied
circumstellar densities in the case of SN\,1988Z are $\sim 10^7$
cm$^{-3}$ (Fabian \& Terlevich 1996).

On the basis of the X--ray properties alone it is difficult to
distinguish between an ultraluminous SN and an AGN for the nuclear emission,
as sufficiently little is
currently known about the X--ray spectral properties of objects
like SN\,1988Z and SN\,1986J (Stevens, Strickland \& Wills 1999). 
Although it is possible that an X--ray luminous SN could account
for part of the extended emission, 
we consider it unlikely to be the main source of nuclear emission, as 
the flat radio spectrum of TH2 and the lack of X--ray 
variability argue against
an ultraluminous SN at the core of NGC 253.   

\subsection{The Infrared Maximum}

The brightest discrete source at [SIII]9532, IR and HCN wavelengths 
appear to be
spatially coincident (indeed we have assumed this for the tables and
figures). Pina \etal (1993) called this source IRS1, while Kalas \&
Wynn--Williams (1994) called it peak 1. We will refer to it as the IR
maximum. It is located at $\alpha$ = 00$^{h}$47$^{m}$33.006$^{s}$, 
$\delta$ = --25$^{\circ}$17$^{'}$18.23$^{''}$ (J2000). It is the strongest
Br$\gamma$ source in the central region with a 2${''}$ aperture flux of 3.2
$\times$ 10$^{-14}$ erg s$^{-1}$ cm$^{-2}$ (Forbes \etal 1993). 
This corresponds to 2 $\times$ 10$^{51}$ ionising photons per sec, or 
about 1000 O stars, and suggests that the IR maximum 
is a powerful 
young star cluster of total mass $\sim$ 10$^5$ (for a normal IMF). 
From the H$\alpha$ emission and a burst model, Watson
\etal (1996) also favour a young ($<$ 100 Myr old) star cluster.
Our measurements of the [SIII] to H$\alpha$ ratio for this source 
(see Table 3) are more
consistent with photoionisation from young stars rather than SNR--produced 
shocks. 

Assuming that the Br$\gamma$ emission comes solely from
HII regions, and an electron temperature of 10$^4$ K, then we would predict a
thermal 
2cm flux density of $\sim$ 3 mJy. If A$_V$ = 5 towards this source, then
the expected flux rises to $\sim$ 5 mJy. The IR maximum does not appear to
have an associated radio source of this strength, indicating a 2cm flux
limit of $\le$ 0.4 mJy. 
So although we favour the
interpretation of a large cluster of young stars, it is
curious that there is no strong radio emission from the associated HII 
regions. 
Paglione \etal (1995) inferred a density of $\sim$ 10$^5$ cm$^{-3}$ 
towards this source which is even higher than that inferred for the
nucleus. Perhaps this very high density gas can act to modify the
Galactic (i.e. low density) relationship between ionising photons and
thermal radio emission.

\subsection{The Ring Region}

The [SIII], H$\alpha$ and IR sources resemble a ring--like structure
surrounding the nucleus (see Fig. 4). It is about 50 pc in diameter.  
As shown by Sams \etal (1994) the appearance of discrete sources is largely
due to the variable dust obscuration within this region. Nevertheless the
presence of strong H$\alpha$ and IR emission in a ring--like structure
suggests that NGC\,253 contains a $\sim$ 50 pc sized starburst ring (as it is
difficult to imagine how dust would create the impression of a ring when it
doesn't exist). 

This ring has also been seen in the form of a molecular torus (Israel,
White \& Bass 1995) with the IR maximum defining the SW edge. The velocity
field around the ring is consistent with solid body rotation (Paglione
\etal 1995). HI observations reveal evidence for a rapidly rotating ring of
cold gas (Koribalski \etal 1995). The optical/IR ring appears to define the
inner edge of a larger ($\sim$ 500 pc sized) cold gas torus, which has a
higher gas density than the outer parts of the torus (Israel \etal 1995). 
Arnaboldi \etal (1995) claim that NGC\,253 has two 
inner Lindblad resonances, one at $\sim$ 300 pc (i.e. within the torus) and
another at about 50 pc radius (i.e. the starburst ring).  

As can be seen in Fig. 4 there is very little spatial correspondence
between the optical/IR sources and the radio sources. A radio spectral
index map does however show that the NW ridge of the ring (which has
the most intense optical/IR sources) has a flat or even positive spectral
index indicating that the radio emission is thermal in nature from HII
regions. The fact that it is diffuse rather than in the form of compact
sources suggests that either the HII regions have overlapped and 
expanded to fill a large area 
or perhaps the emission is being absorbed by high density gas (which may in
turn be associated with the molecular torus).  
The Br$\gamma$ emission is also strongest on the NW ridge of the ring,
while the [FeII] emission traces SNRs 
on the SE ridge of the ring (Forbes \etal
1992). In other words the contribution of HII regions relative 
to SNRs is high on
the NW side of the ring and low on the SE side of the ring. This may
indicate a temporal evolution, with star formation starting on the SE side
and progressing to the NW side of the ring. 

Further support for the starburst interpretation of the ring sources
comes from optical line ratios. In Fig. 5 we show the
diagnostic diagram adapted from Kirhakos \& Phillips (1989) which plots
[SII]6716,6731/H$\alpha$ vs [SIII]9069,9532/H$\alpha$. 
The [SII]6716,6731/H$\alpha$ ratio, which is largely extinction
independent, 
comes from the central region (60 pc)
measurement of Schulz \& Wegner (1992). This therefore 
represents an average over the
central region which includes the ring structure.  
We have directly measured the
[SIII]9532 and H$\alpha$ fluxes for each discrete 
source from our WFPC2 images, and
assuming a ratio of 9532/9069 = 2.5 (Osterbrock \etal 1990) we can estimate
the dereddened ratio for A$_V$ = 5 and 10. We find that the discrete 
sources in the ring are consistent with 
the photoionisation models for OB stars and HII regions, rather than 
the shock models for SNRs. 

In summary, NGC\,253 contains an almost circular, starburst ring of diameter
$\sim$ 50 pc which defines the inner edge of a 500 pc sized torus of cold
gas. The discrete sources in the ring do not, in general, have
associated compact radio sources.
Their appearance at optical, and to some extent IR 
wavelengths, is no doubt affected by dust obscuration but this is unlikely
to artificially create a ring--like structure. 
Gas rings, of pc to kpc sizes, have been seen in a number of ultraluminous
starburst galaxies (Downes \& Solomon 1998). This suggests that 
similar processes to those seen in
NGC\,253 are occurring in other distant galaxies 
but on a much grander scale.


\subsection{The Region Exterior to the Ring}

Outside of the starburst ring there is limited IR coverage in the literature,
however there are additional HCN (Paglione \etal 1995), radio (Ulvestad \&
Antonucci 1997), [SIII] and H$\alpha$ discrete sources. 
As can be seen in Fig. 6, most of these sources, and the
radio sources interior to the ring (Fig. 4), 
are located in a SW--NE direction. The
position angle of these sources is about 40$^o$. 
 
Recently, Peng \etal (1996) has provided an explanation for this
alignment. They suggest that the gas orbits in the central 100 pc 
of NGC\,253 are moving in a potential associated with the galaxy's large
scale bar which has a P.A. $\sim$ 70$^o$ 
(Forbes \& DePoy 1992). At the inclination of NGC\,253, 
these orbits have an apparent position angle on the sky which is similar to
that observed. Gas tends to accumulate at the apocentres of these orbits
leading to clumps of star formation.  


\subsection{Relationship to the Superwind}

The bulk of the X--ray emission from NGC 253 is due to the outflowing
superwind (Read \etal 1997). Its diffuse emission extends about 12$^{'}$ (9
kpc) from the galaxy disk (see Fig. 2). 
We find that the diffuse emission in the 
central 500 pc has a position angle of $\sim$
140$^{\circ}$, i.e. the same as that on large scales and perpendicular to
the galaxy disk. 
Thus the superwind emission extends from a few hundred parsecs to about 
ten thousand parsecs from the galaxy core. 
The central source size (i.e. $\sim$ 120 pc)  is larger
than the starburst ring but well within the HI torus. High resolution
imaging by \axaf\ will confirm whether the torus is acting to collimate the
outflowing superwind, and whether the bulk of its central emission 
originates from the starburst ring or the galaxy nucleus itself. 

\section{Summary and Future Work}

By spatially aligning the discrete sources seen at various wavelengths, we
have explored the nature of these sources in the central regions in NGC 253. 
We find that the optical, IR and mm sources trace out a nearly circular
starburst ring of size $\sim$ 50 pc. There is 
some evidence for the star formation starting on the SE side of the ring. 
The brightest IR source in the central region lies in the SW side of the
ring, and is probably a massive young star cluster.  
This ring
defines the inner edge of a larger ($\sim$ 500 pc) cold gas torus, which
has probably been created by gas inflow from a larger scale bar and
deposited at the inner Lindblad resonance. 
The interior of the ring
contains most of the compact radio sources (which are mostly SNRs) and the
radio nucleus itself. The radio nucleus is almost certainly a heavily
obscured AGN, which is not visible at optical or IR wavelengths and
contributes only some fraction of the X--ray emission from the central
10$^{''}$ (120 pc). 
The diffuse emission associated with the outflowing superwind is present in
the central region, on a size scale consistent with the idea of collimation
by the gas torus.

We have already mentioned that \axaf\ will be important in probing the
true nature of the X--ray emission in the nuclear region. 
It will also be useful to search for new SN. 
From the Ulvestad \& Antonucci (1997) list of radio sources, there are 4
spatially resolved SNRs. Their  
average size is $\sim$ 0.1$^{''}$ or 1.2 pc in diameter. A SNR expanding at
a rate of 10$^4$ km s$^{-1}$ will reach this size in about 60 yrs. 
Ulvestad \& Antonucci have recalculated the SN rate in NGC\,253 based on the
fading of radio sources (i.e. no more than 2\% yr$^{-1}$) and the lack of new
sources over an 8 year baseline. They conclude that the rate is about 0.1
SN yr$^{-1}$ with an upper limit of 0.3 SN yr$^{-1}$. Therefore extending
the time baseline by a few more years should be a fruitful exercise.
The increased sensitivity and spatial resolution of \axaf\ will be ideal to
search for the X--ray emission from recent SN, particularly in the 
dusty circumnuclear region.

\section*{Acknowledgements}

We thank B. Koribalski and the referee
(T. Heckman) for some helpful comments, 
Support for this work was provided by NASA through grant number
GO-06440.01-95A from the Space Telescope Science Institute, which
is operated by the Association of Universities 
for Research in Astronomy, Inc., under NASA contract NAS5-26555.

\newpage

\phantom{a}
{\bf Table 1.}{\it ROSAT} observations of NGC 253.

\medskip
\begin{tabular}{c c c c c }
\hline
Instrument & Exposure (ks) & ROR \# & P.I. & Start date\\
\hline

PSPC & 11.6 & rp600087a00 & Pietsch & 1991.359\\ 
PSPC & 11.2 & rp600087a01 & Pietsch & 1992.155\\
HRI & 25.7 & rh600088a01 & Pietsch & 1992.157\\
HRI & 11.0 & rh600714n00 & Pietsch & 1995.003\\
HRI & 19.8 & rh600714a01 & Pietsch & 1995.164\\

\hline

\end{tabular}

\phantom{a}
\vspace{0.1in}
\noindent
Notes: Columns list the {\it ROSAT} instrument, exposure time, ROSAT filename,
principal investigator and observing start date.\\

\newpage

\phantom{a}
{\bf Table 2.} Optical Sources.

\medskip
\begin{tabular}{c c c c c}
\hline

Position (J2000) & Offset from & [SIII]9532 flux 
& H$\alpha$ flux & Log([SIII]/H$\alpha$)\\

$\alpha$=00$^{h}$47$^{m}$   $\delta$=--25$^{\circ}$17$^{'}$ & Nucleus  
& $\times$10$^{-15}$ & $\times$10$^{-15}$ & A$_{V}$=10\ \ \ \ \ A$_{V}$=5\\

& (arcsec) & (erg cm$^{-2}$s$^{-1}$) & (erg cm$^{-2}$s$^{-1}$) &  \\

\hline
33.01$^{s}$\ \ \ 18.2$^{''}$ & --2.21, --1.17 &  27.1  & 1.5  & \ 0.5\ \ \ \ \ \ \ \ \ \ 1.0 \\
33.02$^{s}$\ \ \ 18.8$^{''}$ & --1.98, --1.70 &  10.5  & 4.5  & --0.4\ \ \ \ \ \ \ \ \ \ 0.1 \\
33.06$^{s}$\ \ \ 17.1$^{''}$ & --1.54, --0.03 &  4.6   & 1.1  & --0.1\ \ \ \ \ \ \ \ \ \ 0.3 \\
33.08$^{s}$\ \ \ 17.0$^{''}$ & --1.26, +0.13 &  4.0   & 1.0  & --0.1\ \ \ \ \ \ \ \ \ \ 0.3 \\
33.10$^{s}$\ \ \ 16.7$^{''}$ & --0.91, +0.40 &  4.4   & 0.4  & \ 0.4\ \ \ \ \ \ \ \ \ \ 0.8 \\
33.11$^{s}$\ \ \ 16.1$^{''}$ & --0.83, +0.94 &  4.2   & 0.3  & \ 0.4\ \ \ \ \ \ \ \ \ \ 0.9 \\
33.14$^{s}$\ \ \ 18.1$^{''}$ & --0.38, --1.06 &  3.3   & 0.8  & --0.1\ \ \ \ \ \ \ \ \ \ 0.3 \\
33.18$^{s}$\ \ \ 18.0$^{''}$ &   +0.15, --0.98 &  2.5   & 0.5  & \ 0.0\ \ \ \ \ \ \ \ \ \ 0.4 \\
33.19$^{s}$\ \ \ 16.2$^{''}$ &   +0.22, +0.97 &  3.1   & 1.4  & --0.4\ \ \ \ \ \ \ \ \ \ 0.1 \\
33.24$^{s}$\ \ \ 17.0$^{''}$ &   +1.00, +0.15 &  2.8   & 0.9  & --0.2\ \ \ \ \ \ \ \ \ \ 0.2 \\
\medskip
33.24$^{s}$\ \ \ 16.5$^{''}$ &   +1.06, +0.71 &  4.4   & 1.8  & --0.3\ \ \ \ \ \ \ \ \ \ 0.1 \\
33.27$^{s}$\ \ \ 14.2$^{''}$ &   +1.42, +2.91 &  2.2   & 1.0  & --0.4\ \ \ \ \ \ \ \ \ \ 0.1 \\
33.45$^{s}$\ \ \ 12.7$^{''}$ &   +3.81, +4.47 &  6.8   & 0.8  & \ 0.2\ \ \ \ \ \ \ \ \ \ 0.7 \\
33.83$^{s}$\ \ \ 10.7$^{''}$ &   +8.90, +6.38 &  1.4   & 0.3  & --0.1\ \ \ \ \ \ \ \ \ \ 0.4 \\
34.08$^{s}$\ \ \ 08.5$^{''}$ &  +12.47, +8.72 &  1.9   & 0.3  & \ 0.1\ \ \ \ \ \ \ \ \ \ 0.6 \\
\hline

\end{tabular}

\phantom{a}
\vspace{0.1in}
\noindent
Notes: Positions of the discrete sources have been shifted so that the
brightest [SIII]9532 source lies at $\alpha$ = 00$^{h}$47$^{m}$33.006$^{s}$,
$\delta$ = --25$^{\circ}$17$^{'}$18.23$^{''}$ (J2000). 
Offset is the position of the discrete source relative to the radio
nucleus, ie $\alpha$=00$^{h}$47$^{m}$33.169$^{s}$,
$\delta$=--25$^{\circ}$17$^{'}$17.06$^{''}$ (J2000).  
The [SIII]9532 and H$\alpha$ fluxes are taken from a 
0.4$^{''}$ diameter aperture centred on the discrete source maximum.  
The [SIII]9069,9532/H$\alpha$ ratio is calculated from [SIII]9532/H$\alpha$ using the relation, 
9532/9069 = 2.5 (Osterbrock {\it et al.} 1990) and dereddened for 
extinctions of A$_{V}$=10 and A$_{V}$=5. Sources within the ring
are listed first, and then 4 sources that lie beyond the ring.\\

\newpage

\phantom{a}
{\bf Table 3.} Infrared/mm Sources.

\medskip
\begin{tabular}{c c c c }
\hline
Position (J2000)& \ \ \ Offset\ \ \ & Wavelength & Reference\\
$\alpha$=00$^{h}$47$^{m}$ \ $\delta$=--25$^{\circ}$17$^{'}$ &(arcsec) &  & \\
\hline
32.901$^{s}$\ \ \ 19.24$^{''}$ & --3.623, --2.18 & 1.65 ${\it\mu}$m & 1\\
33.006$^{s}$\ \ \ 18.23$^{''}$ & --2.214, --1.17 & 1.65 ${\it\mu}$m & 1\\
33.062$^{s}$\ \ \ 17.13$^{''}$ & --1.456, 0.00 & 1.65 ${\it\mu}$m & 1\\
33.161$^{s}$\ \ \ 18.05$^{''}$ & --0.104, --0.99 & 1.65 ${\it\mu}$m & 1\\
33.241$^{s}$\ \ \ 16.40$^{''}$ & +0.981, +0.66 & 1.65 ${\it\mu}$m & 1\\
\hline
33.006$^{s}$ \ \ \ 18.23$^{''}$ & --2.214, --1.17  & 8.5 - 12.5 ${\it\mu}$m & 2\\
33.185$^{s}$ \ \ \ 16.53$^{''}$ & +0.223, +0.53 & 8.5 - 12.5 ${\it\mu}$m & 2\\
\hline
33.006$^{s}$ \ \ \  18.23$^{''}$ & --2.214, --1.17 & 10 - 20 ${\it\mu}$m &3\\
33.116$^{s}$ \ \ \  16.63$^{''}$ & --0.724, +0.43  & 10 - 20 ${\it\mu}$m &3\\
\hline
33.006$^{s}$ \ \ \  18.23$^{''}$ & --2.214, --1.17  & 1.6 - 4.8 ${\it\mu}$m & 4\\
33.156$^{s}$ \ \ \  16.39$^{''}$ & --0.183, +0.67 & 4.8 ${\it\mu}$m &4\\
33.228$^{s}$ \ \ \  16.51$^{''}$ &  +0.805, +0.55  & 1.6 - 3.38 ${\it\mu}$m &4\\
33.450$^{s}$ \ \ \  12.79$^{''}$ &  +3.813, +4.27 & 1.6 - 4.8 ${\it\mu}$m &4\\
33.829$^{s}$ \ \ \  10.74$^{''}$ &  +8.957, +6.32 & 1.6 - 3.38 ${\it\mu}$m &4\\
\hline
32.283$^{s}$\ \ \ 16.82$^{''}$  & --12.010, +0.24  & HCN J=1 $\rightarrow$ 0 &  5\\
33.006$^{s}$\ \ \ 18.23$^{''}$  & --2.214, --1.17  & HCN J=1 $\rightarrow$ 0 &  5\\
33.162$^{s}$\ \ \ 15.43$^{''}$  & --0.100, +1.63  & HCN J=1 $\rightarrow$ 0 &  5\\
33.316$^{s}$\ \ \ 14.73$^{''}$  & +1.998, +2.33  & HCN J=1 $\rightarrow$ 0 &  5\\
33.368$^{s}$\ \ \ 13.33$^{''}$  & +2.703, +3.73  & HCN J=1 $\rightarrow$ 0 &  5\\
33.626$^{s}$\ \ \ 11.94$^{''}$  & +6.196, +5.12  & HCN J=1 $\rightarrow$ 0 &  5\\
33.781$^{s}$\ \ \ 11.24$^{''}$  & +8.294, +5.82  & HCN J=1 $\rightarrow$ 0 &  5\\
34.091$^{s}$\ \ \ 10.54$^{''}$  & +12.503, +6.52  & HCN J=1 $\rightarrow$ 0 &  5\\
\hline
\end{tabular}

\phantom{a}

\noindent
Notes: Positions of the discrete sources have been shifted so that the
brightest IR and mm sources lie at $\alpha$ = 00$^{h}$47$^{m}$33.006$^{s}$,
$\delta$ = --25$^{\circ}$17$^{'}$18.23$^{''}$ (J2000). Offsets are measured
relative to the radio nucleus, ie $\alpha$ = 00$^{h}$47$^{m}$33.169$^{s}$,
$\delta$ = --25$^{\circ}$17$^{'}$17.06$^{''}$ (J2000). The references for
the IR and mm observations are: 
(1) = Forbes, Ward \& DePoy 1991; (2) = Keto et
al. 1993; (3) = Pi\~{n}a {\it et al.} 1992; 
(4) = Kalas \& Wynn-Williams 1994;  (5)
= Paglione {\it et al.} 1995.

\newpage

\phantom{a}
{\bf Table 4.} Radio Sources.
\medskip

\begin{tabular}{c c c c c c}

\hline
Position (J2000) & Offset & Flux & Luminosity & Spectral Index & TH\\
$\alpha$=00$^{h}$47$^{m}$\ \ \ 
$\delta$=--25$^{\circ}$17$^{'}$ & (arcsec) 
& (mJy) & (10$^{35}$$\ $erg$\ $s$^{-1}$) & 
$\alpha^{6}_{2}$ & Source\\
\hline

32.865$^{s}$\ \ \ 21.29$^{''}$ & --4.123, --4.23 &15.72 &  17.63 
&  --0.69$\pm$0.06 & TH9\\
32.922$^{s}$\ \ \ 20.20$^{''}$ & --3.350, --3.14 & 4.44 
&  4.98  &  +0.17$\pm$0.10 &\\
32.974$^{s}$\ \ \ 19.64$^{''}$ & --2.645, --2.58 & 4.09 &  4.59  & &\\
33.001$^{s}$\ \ \ 19.34$^{''}$ & --2.278, --2.28 & 5.14 &  5.76  
&  --0.24$\pm$0.07 & TH7\\
33.039$^{s}$\ \ \ 18.75$^{''}$ & --1.763, --1.69 & 0.96 &  1.08  & &\\


33.104$^{s}$\ \ \ 18.08$^{''}$ & --0.882, --1.02 & 7.82 &  8.77  
&  +0.01$\pm$0.07 & TH6\\
33.105$^{s}$\ \ \ 17.55$^{''}$ & --0.868, --0.49 & 3.39 &  3.80  & &\\
33.129$^{s}$\ \ \ 19.85$^{''}$ & --0.542, --2.79 & 3.51 &  3.94  
&  --0.71$\pm$0.09 & \\
33.139$^{s}$\ \ \ 17.68$^{''}$ & --0.407, --0.62 & 1.80 &  2.02  & & TH5\\


33.157$^{s}$\ \ \ 17.39$^{''}$ & --0.163, --0.33 & 13.08 &  14.67 & & TH4\\
33.165$^{s}$\ \ \ 17.77$^{''}$ & --0.054, --0.71 & 4.05 &  4.54  & & TH3\\
33.169$^{s}$\ \ \ 17.06$^{''}$ & 0.000, 0.00 & 37.14 &  41.64 
&  +0.04$\pm$0.06 & TH2\\
33.173$^{s}$\ \ \ 19.05$^{''}$ & +0.149, +0.20 & 1.49 &  1.67  
&  --0.79$\pm$0.12 &\\
33.180$^{s}$\ \ \ 16.86$^{''}$ & +0.054, --1.99 & 2.99 &  3.35  & &\\
33.284$^{s}$\ \ \ 15.47$^{''}$ & +1.560, +1.59 & 5.59 &  6.27  
&  +0.35$\pm$0.08 & TH1\\
33.375$^{s}$\ \ \ 15.01$^{''}$ & +2.794, +2.05 & 1.91 
&  2.14  &  --0.78$\pm$0.10 & \\


\hline

\end{tabular} 

\phantom{a}

\noindent
Notes: The table lists radio sources with 2cm luminosity 
$>$10$^{35}$ erg s$^{-1}$ (from Ulvestad \& Antonucci 1997). 
Offset is from the radio nucleus, ie 
$\alpha$ = 00$^{h}$47$^{m}$33.169$^{s}$, $\delta$ =
--25$^{\circ}$17$^{'}$17.06$^{''}$ (J2000). 
The spectral index is from 6 to 2cm. The TH source is the identification
number from Turner \& Ho (1985).

\newpage

\begin{figure}
\vspace*{12cm}
\includegraphics{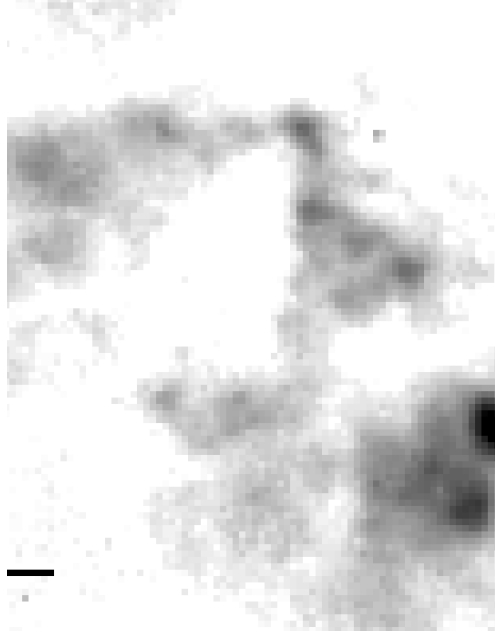}
\caption{ 
A [SIII]9532 emission line image of the 
central $\sim$ 50pc of NGC\,253. North is up and East is left. 
The horizontal line represents 1$^{''}$ (12 pc). The image
shows a collection of discrete [SIII] line emission sources.
The brightest [SIII] line emission source is also the location of
the IR maxiumum, which defines the SW edge of a ring of emission sources.
}
\label{fig1}
\end{figure}

\begin{figure}
\vspace*{12cm}
\includegraphics{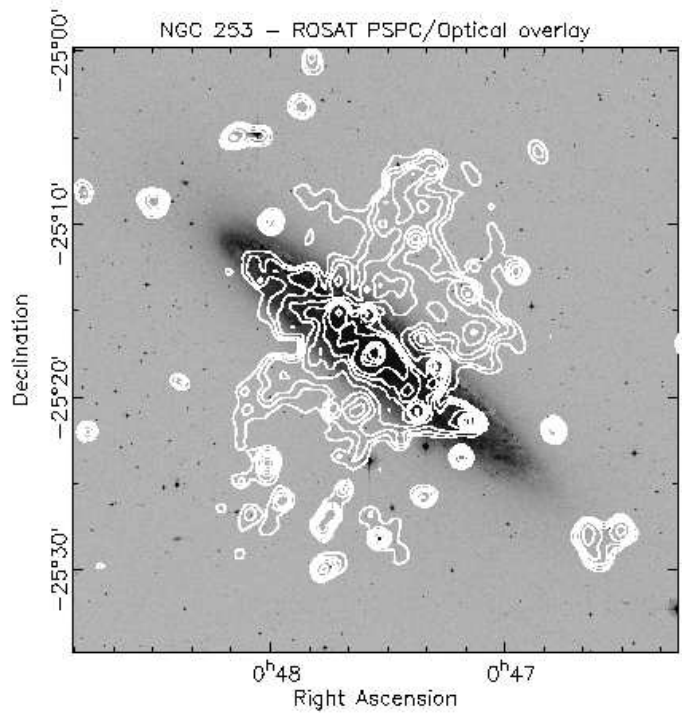}
\caption{
\rosat\ PSPC X--ray contours superposed on the Digitised Sky
Survey optical image of NGC\,253. The X--ray emission comes from a variety of
sources; bi--conical emission perpendicular to the disk, diffuse disk
emission and point sources.
}
\label{fig2}
\end{figure}

\begin{figure}
\vspace*{12cm}
\includegraphics{fig3a.ps}
\includegraphics{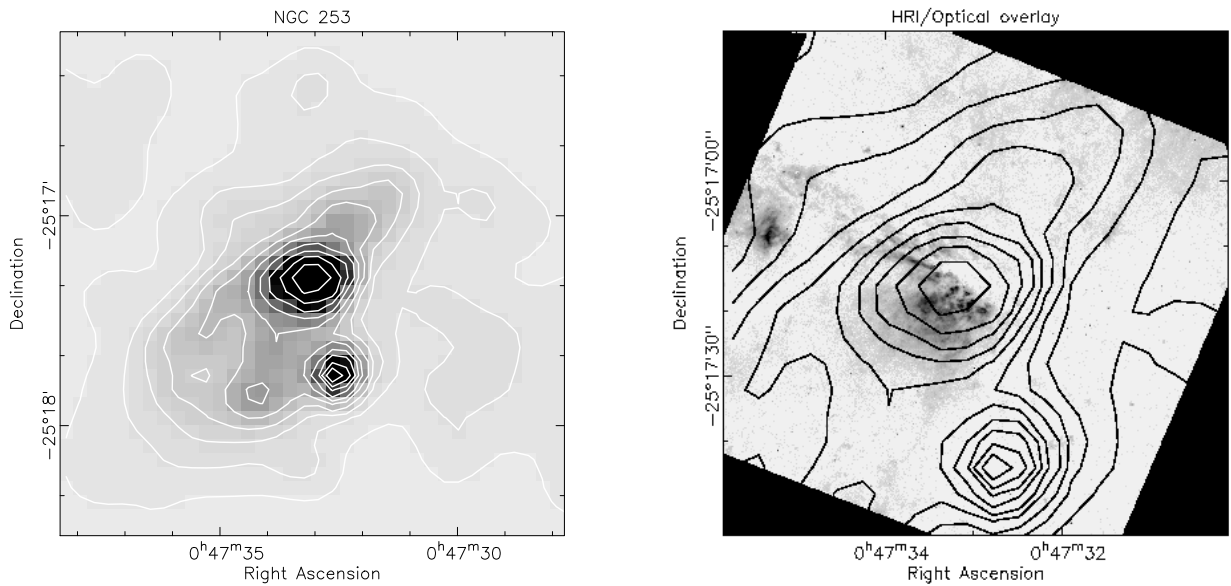}
\caption{
Left: \rosat\ HRI image of the central regions of NGC\,253. An
extended nuclear source is seen along with diffuse emission and a
point--like source $\sim$ 25$^{''}$ South (most likely an X--ray
binary). 
The contours levels start at 1.15 ct s$^{-1}$ arcmin$^{-2}$
and increase by 50\% per contour level. 
Right: \rosat\ HRI X--ray contours superposed on a WFPC2 H$\alpha$
emission line image. The H$\alpha$ filaments are contained within the
bi--conical X--ray contours. The bright X-ray source at R.A. $\sim$ 
00$^h$47$^m$33$^s$, Dec. $\sim$ --25$^{o}$17$^{'}$45$^{''}$ is probably
a foreground source.
} 
\label{fig3}
\end{figure}

\begin{figure}
\vspace*{12cm}
\includegraphics{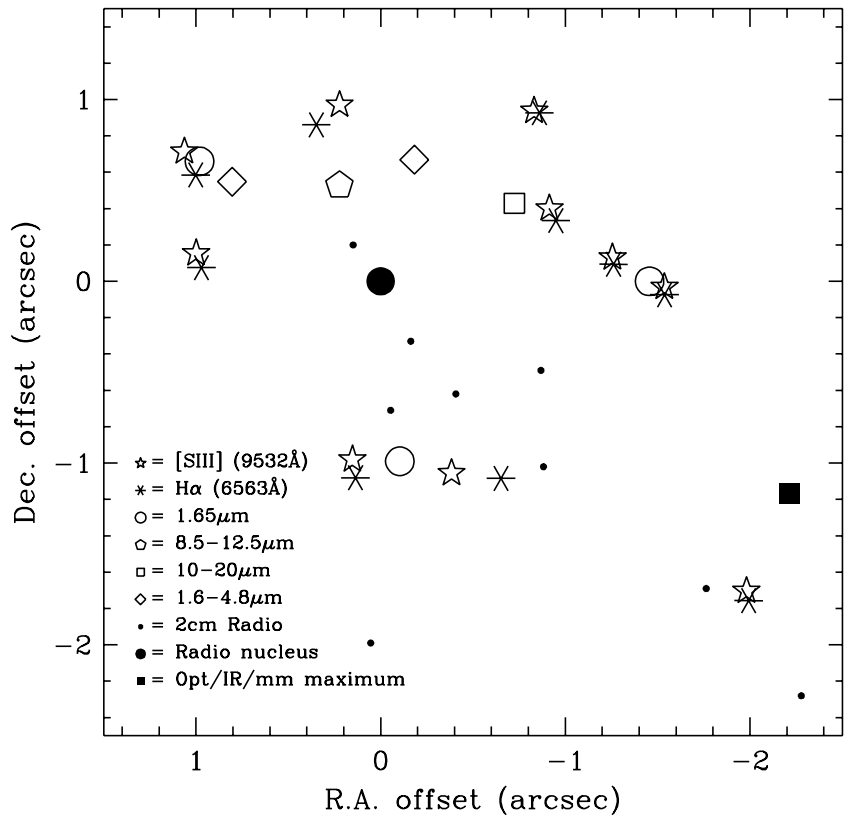}
\caption{
Location of discrete sources in the central region of
NGC\,253. The optical, infrared and millimeter sources reveal a ring--like
structure, whereas the compact 
radio sources lie along a SW--NE line. There is very
little if any spatial correspondence between the radio sources and the
emission at other wavelengths. The radio 
nucleus has no strong optical, infrared
or millimeter counterpart.
}
\label{fig4}
\end{figure}

\begin{figure}
\vspace*{12cm}
\includegraphics{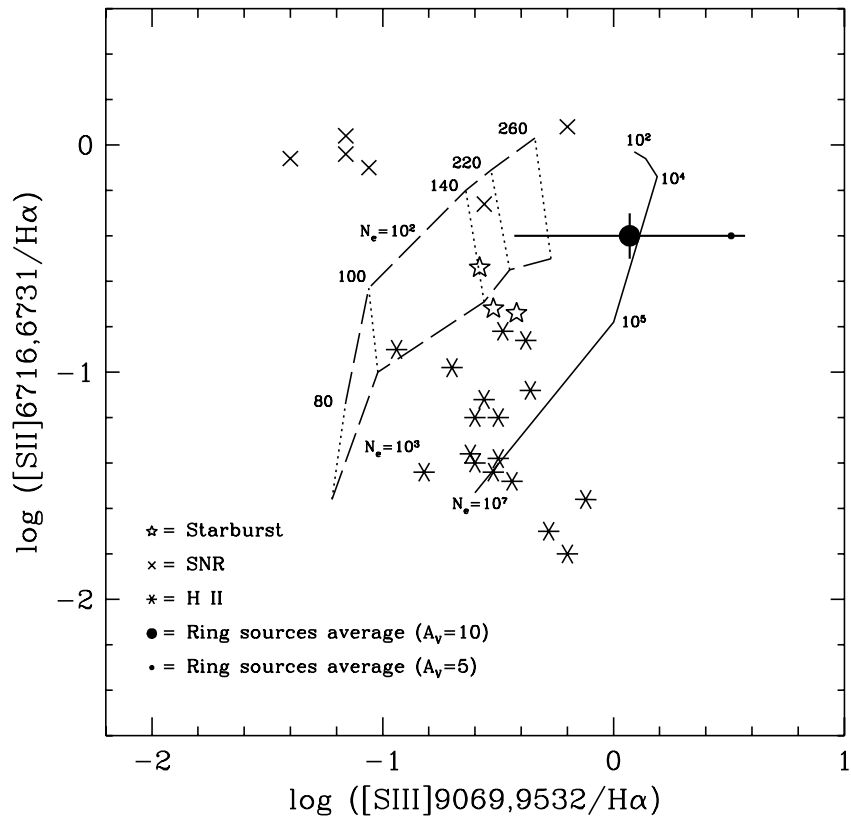}
\caption{
Emission line diagnostic diagram. The solid line represents
a photoionisation model with different electron densities labelled, 
the dashed lines 
represent shock models for a range of shock velocities at two different
densities. The location of typical starburst galaxies, 
Galactic SNRs and HII regions are also shown. The mean and range of values
for the 
central region of NGC\,253, with A$_V$ = 10, 
is shown by a large filled circle with a horizontal line.  
The small filled circle shows the mean value of the line
ratios for A$_V$ = 5. The [SII] and [SIII] emission line ratios are more
consistent with photoionisation by young OB stars than SNR--driven
shocks.
}
\label{fig5}
\end{figure}

\noindent
\begin{figure}
\vspace*{12cm}
\includegraphics{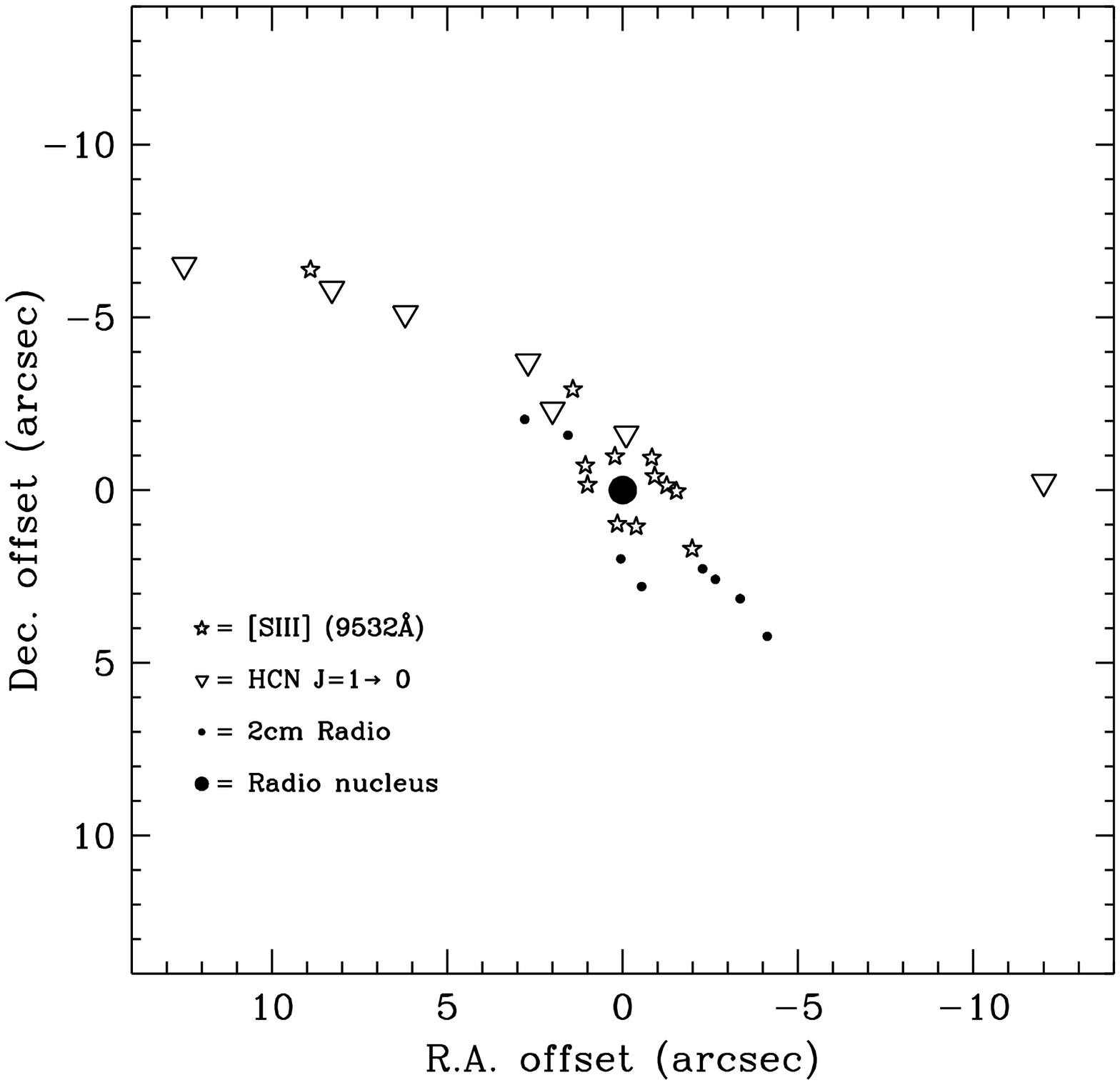}
\caption{
Location of discrete sources beyond the central ring in 
NGC\,253. Most sources lie in a SW--NE direction.
}
\label{fig6}
\end{figure}

\end{document}